# Practical Power System Inertia Monitoring Based on Pumped Storage Hydropower Operation Signature

Hongyu Li, *Member, IEEE*, Chang Chen, *Student Member, IEEE,* Mark Baldwin, Shutang You, *Senior Member, IEEE*, Wenpeng Yu, *Member, IEEE*, Lin Zhu, *Senior Member, IEEE*, Yilu Liu, *Fellow, IEEE*

*Abstract*—**This paper proposes a practical method to monitor power system inertia using Pumped Storage Hydropower (PSH) switching-off events. This approach offers real-time system-level inertia estimation with minimal expenses, no disruption, and the inclusion of behind-the-meter inertia. First, accurate inertia estimation is achieved through improved RoCoF calculation that accounts for pre-event RoCoF, reducing common random frequency fluctuations in practice. Second, PSH field data is analyzed, highlighting the benefits of using switching-off events for grid inertia estimation. Third, an event detection trigger is designed to capture pump switching-off events based on local and system features. Fourth, the method is validated on the U.S. Eastern Interconnection model with over 60,000 buses, demonstrating very high accuracy (3%-5% error rate). Finally, it is applied to the U.S. Western Interconnection, with field validation showing a 9.9% average absolute error rate. Despite challenges in practical power system inertia estimation, this method enhances decision-making for power grid reliability and efficiency, addressing challenges posed by renewable energy integration.**

## I. INTRODUCTION

REAL-TIME inertia estimation is increasingly crucial for power system operations due to the rising penetration of renewable energy sources [1]-[4]. To understand the importance of monitoring inertia, consider a boat (the power system) on a river (energy flow). Traditional heavy cargo (conventional generators) stabilizes the boat, making it resistant to disturbances. Switching to lighter, renewable energy sources is like replacing heavy cargo with lighter materials, increasing sensitivity to disturbances. Without enough weight (inertia), minor waves can destabilize the boat, causing system outages. Power system operator must monitor and manage inertia to maintain system stability, much like a captain adjusting the boat's load. An effective inertia monitoring tool is essential for ensuring stability during this transition.

Various practical methods for estimating inertia in power systems exist, including dispatch-based online generator inertia summation, generator-event-driven methods, probing signal-based methods, and ambient signal-based methods, etc. However, these approaches have limitations that hinder their practical application. The dispatch-based survey method of online generator inertia summation may overlook inertia from IBRs and behind-the-meter resources, such as load inertia and synthetic inertia from renewables [5], [6]. Probing-based methods necessitate additional investments to generate probing signals for the power system [7], [8]. Ambient-based methods rely on the frequency noise of the power system, introducing accuracy concerns due to uncertainty from noise sources [9]-[11]. Event-driven methods utilize sudden Megawatt (MW) change events in the power system, which are considered more accurate in estimating inertia due to encompassing all types of inertia from the frequency response following the event [12]-[17]. However, this method is constrained by insufficient generator trip event numbers and typically relies on offline analysis due to limited confirmed event information.

Besides the above practical methods, there are insightful theoretical approaches to power system inertia estimation. An innovative oscillation-information-based method in [18], [19] establishes a relationship between electromechanical oscillation information (e.g., frequency, damping, eigenvalue) and area inertia. By extracting the information from measurements, the inertia of oscillating units or areas can be estimated. However, units or areas with weak or no oscillations face challenges in inertia estimation, making it difficult to estimate interconnection-level system inertia in practice. AI-based methods in [20] use machine learning techniques like long-recurrent convolutional neural networks and graph convolutional neural networks to achieve high inertia estimation accuracy in simulation system. Its application in real-world power systems is not discussed. Modeling-based inertia estimation methods involve building a frequency response model, measuring frequency data, and extracting inertia. Various models exist for this method, including the Bayesian-approach-based model [21], [22], local rational model approach [23], transfer functions [24], [25], subspace state-space models [26], ARMAX [27], [28], discrete-time linear systems [29], and dynamic regressor and mixing models [30]. While these methods have the advantage of a clear mechanism, they are challenging to apply to actual interconnection-level power systems due to the difficulty in building accurate models for large interconnection systems.

To address these challenges, this paper presents a novel practical method for interconnection-level power system inertia estimation based on pump switching-off events [31]. This method leverages the constant large disturbances of Pumped Storage Hydropower (PSH) events to significantly improve inertia estimation. The PSH switching-off event

This work was supported by Water Innovation for a Resilient Electricity System (HydroWIRES) Initiative from U.S. DOE's Water Power Technologies Office (WPTO) under agreement 39197.

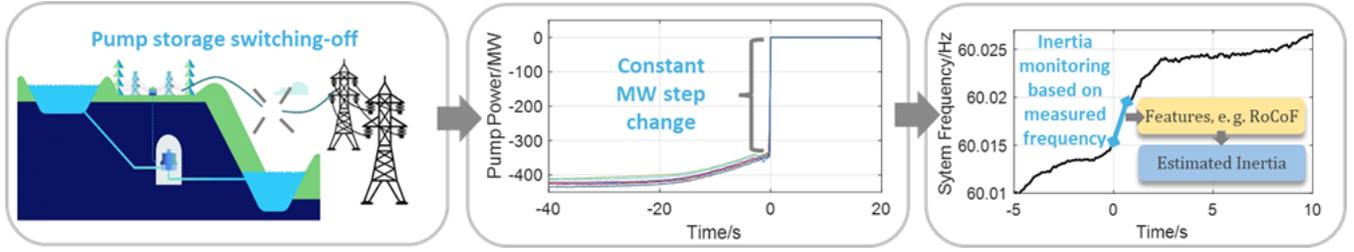

**Fig. 1.** Concept of inertia estimation using PSH operation signatures.

offers the advantage of constant step power change. When the guide vane opening or output power decreases to a predetermined threshold, the breaker deactivates, resulting in a consistent instantaneous power change injected into the grid. This advantage eliminates the need for real-time power measurements, if historical or manufacturer's setting data of pump switching-off power is available.

The high-level concept of the proposed inertia monitoring method is illustrated in Fig. 1. FNET/GridEye frequency monitors developed by Oak Ridge National Laboratory (ORNL) and The University of Tennessee Knoxville (UTK) are deployed across the U.S. to capture high-resolution frequency data. An event trigger detects pump switching-off events. High-accuracy Rate of Change of Frequency (RoCoF) is calculated, and then system inertia is estimated real-time.

The detailed contributions of this paper are as follows:

1) Proposing a novel inertia monitoring method based on PSH switching-off events and analyzing field data to highlight its practical benefits.

2) Developing an accurate inertia estimation algorithm using improved RoCoF, which excludes pre-fault RoCoF from widespread random frequency fluctuations in practice.

3) Designing an online event trigger to capture real-time PSH switching-off events, utilizing local features like frequency spikes and voltage steps for high performance.

4) Validating the proposed method using the U.S. EI model and implementing it in the U.S. Western Interconnection, with both simulation and field results showing high accuracy.

This proposed practical method offers several benefits, including real-time, system-level, high-accuracy inertia estimation with no disruptive effects, minimal additional costs, and consideration of behind-the-meter inertia, etc. Details of several specific advantages are:

1) Non-Intrusive: The method uses existing PSH signatures, eliminating the need for additional expensive power modulators and avoiding intrusive synthetic signals. This makes it more advantageous than probing-signal-based methods in [7], [8].

2) Cost-Effective: The method does not require real-time power measurements from expensive Phasor Measurement Units (PMUs). The constant power step signal from the pump switching-off process can be obtained from offline confirmation using public pump unit capacity or historical data. Additionally, it is compatible with low-cost sensors for frequency measurements, such as the Frequency Disturbance Recorder (FDR) from ORNL and UTK. This is more cost-effective than traditional methods based on generator trips [12]-[17] and ambient-based methods [9]-[11] that rely on costly PMUs.

3) High-Accuracy: The method enhances accuracy by monitoring all system inertia through measured frequency response, which reflects all inertia influences after the pump switching-off. Also, the improved RoCoF calculation reduces the impact of background random frequency fluctuations. This is more accurate than the dispatch-based survey method of online generator inertia summation, which often overlooks load inertia and synthetic inertia from renewables [5], [6].

The remaining paper is structured as follows: Section II introduces the inertia estimation algorithm based on improved RoCoF calculation. Section III provides a comprehensive analysis of PSH event field data. Section IV develops an online event detection trigger to capture pump switching-off events. Section V validates the proposed method on the U.S. EI model. Section VI discusses the field application in the U.S. Western Interconnection. Section VII concludes the paper.

## II. EVENT-DRIVEN INERTIA ESTIMATION ALGORITHM USING IMPROVED RoCoF CALCULATION FOR ACTUAL POWER SYSTEM MEASUREMENTS

The event-driven method usually estimates inertia using the RoCoF during an event, combined with the confirmed power change. These events typically involve significant MW step-change disturbances, such as unscheduled generation trips or sudden major load changes. Due to background random frequency fluctuations in actual power system frequency measurements, an improved RoCoF calculation is necessary to minimize these impacts to ensure practical utilization.

### A. Inertia Estimation Algorithm Based on RoCoF

A disturbance with a sudden MW change in power can disrupt the power balance in a system and thus bring frequency response. The initial frequency response following the event is mainly determined by system inertia. Inertia is a critical measure that reflects the system's ability to withstand and recover from such disturbances. Generally, a system with higher inertia will have a slower frequency response, indicating greater resilience to disturbances. This relationship between inertia, power imbalances, and frequency is represented by the swing equation, as outlined below [32], [33]:

$$\frac{2H}{60}\frac{\mathrm{d}f}{\mathrm{d}t} = \Delta P \tag{1}$$

where $H$ is the system inertia; $f$ denotes the system frequency after the event. $\Delta P$ is the sudden event size, which could be from a generator loss or load shedding.

Based on (1), the inertia of the power system can be estimated using the following equation

$$H = \frac{60 \cdot \Delta P}{2 \cdot RoCoF} \qquad (2)$$

Specifically, by calculating RoCoF and verifying the power change caused by an event, the event-driven method can estimate the system's inertia. Generally, this method can be readily implemented using existing measurement equipment and infrastructure, offering a cost-effective and straightforward approach for inertia estimation in power systems.

### B. High-Accuracy RoCoF Calculation to Mitigate Random Frequency Fluctuation Impacts

Due to random MW changes from load and generation in actual power systems, frequency measurements always experience random fluctuations, especially in large interconnected systems. Fig. 2 illustrates an example of random frequency fluctuations.

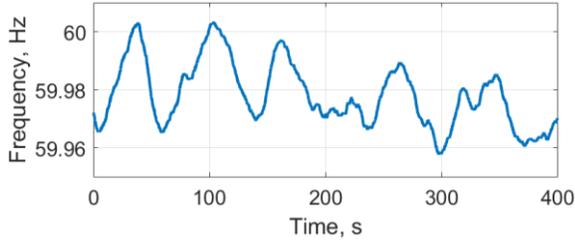

**Fig. 2.** Background random frequency fluctuations from measurements during periods with no events occurring.

During an event, these random frequency fluctuations can affect the true event RoCoF by introducing pre-event RoCoF. An example of background random frequency fluctuation causing pre-event RoCoF is shown in Fig. 3. In this case, due to the background frequency fluctuation, there is a pre-event RoCoF of 5.4 mHz/s. The post-event RoCoF is 13.8 mHz/s, resulting in a net true event RoCoF of approximately 8.4 mHz/s. If the traditional method uses only the post-fault RoCoF of 13.8 mHz/s as the event RoCoF for inertia estimation, the estimated inertia could have a significant error ($=5.4/13.8 \approx 39.13\%$) due to impact of pre-event RoCoF from random frequency fluctuations. Thus, to obtain an accurate event RoCoF and eliminate the influence of random frequency fluctuations, the ture event RoCoF is defined as the post-event RoCoF minus the pre-event RoCoF, as shown in (3). This approach provides the actual event RoCoF by removing the impact of random frequency fluctuations, ensuring a more precise inertia estimation in pratice.

*True Event RoCoF=Post-event RoCoF - Pre-event RoCoF* (3)

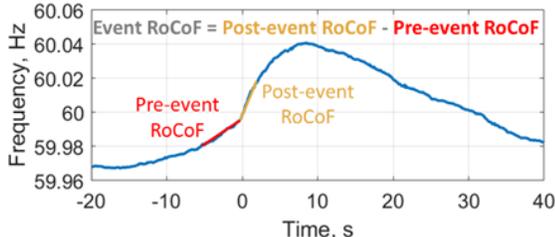

**Fig. 3.** Improved event RoCoF by eliminating impact of pre-event RoCoF due to background random frequency change.

Implementing this improved RoCoF calculation enhances the accuracy of inertia estimation, enabling operators to make better-informed decisions. This RoCoF-based method also reduces reliance on costly and complex measurement infrastructure, making it a practical solution for real-world applications. The following sections will further demonstrate the effectiveness of this approach through various validation scenarios and field applications.

## III. PUMPED STORAGE HYDROPOWER EVENT ANALYSIS USING FIELD DATA

Inertia estimation using the event-driven method requires a substantial number of MW change events. Generator trip events are not ideal for online inertia estimation due to their limited number and unscheduled nature. Operational events from PSH plants offer a better alternative, providing a significant number of events for inertia estimation without compromising system stability. We conducted a comprehensive analysis of real-world PSH data to evaluate its potential for inertia estimation [34]. The analysis utilizes PMU data from the Bath County PSH plant and Raccoon Mountain PSH plant in the United States. Additionally, FNET/GridEye frequency data are incorporated into this research. Findings from this investigation provide valuable insights into leveraging pump switching-off for accurate inertia estimation.

### A. Overview of Pumped Storage Hydropower Plants

PSH plants play a critical role in integrating intermittent renewable sources such as wind and solar into the power grid. They operate by pumping water from a lower reservoir to a higher one during periods of low demand, then releasing it to generate electricity during peak demand. This process effectively manages supply and demand fluctuations, enhancing grid stability by providing valuable inertia. Thus, PSH plants have seen significant deployment globally. Their widespread presence indicates the feasibility of the proposed method for inertia monitoring in most power grids worldwide.

The United States has developed numerous PSH plants. Those with pump unit capacities exceeding 300 MW are listed in Table I. PSH plants with larger unit capacities are more desirable for inertia estimation because they generate pump switching-off events of higher magnitude, resulting in frequency responses with higher magnitude brings higher Signal-to-Noise-Ratios (SNRs), when considering background frequency noise. In the U.S., the Bath County (Eastern Interconnection, EI), Raccoon Mountain (EI), and Helms (Western Electricity Coordinating Council, WECC) PSH plants have the largest pump unit capacities.

TABLE I
U.S. PUMPED STORAGE HYDROPOWER PLANTS WITH UNIT CAPACITIES LARGER THAN 300MW

| PSH plants | Region | Unit capacity (MW) | Unit number |
|---|---|---|---|
| Bath County | U.S. EI | 477 | 6 |
| Raccoon Mountain | U.S. EI | 428.4 | 4 |
| Helms | U.S. WECC | 351 | 3 |
| Ludington | U.S. EI | 329.8 | 6 |
| Bad Creek | U.S. EI | 324 | 4 |
| Bear Swamp | U.S. EI | 300 | 2 |

## B. Four Operational Event Types in Pumped Storage Hydropower Plants

PSH plants typically have four types of switching events: generator switching-off, generator switching-on, pump switching-off, and pump switching-on, as depicted in Fig. 4. PMU power curves for these event types are shown in Fig. 5. Clearly, the pump switching-off process exhibits a distinct step power change, making it suitable for event-based inertia estimation. In contrast, pump switching-on, generator switching-off, and generator switching-on events lack a clear power step change, rendering them less suitable for RoCoF-based inertia estimation.

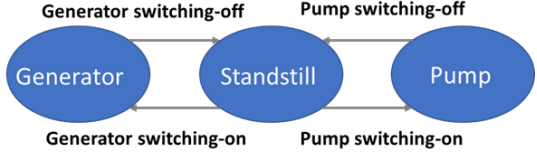

**Fig. 4.** Four types of events in pumped storage hydropower plants.

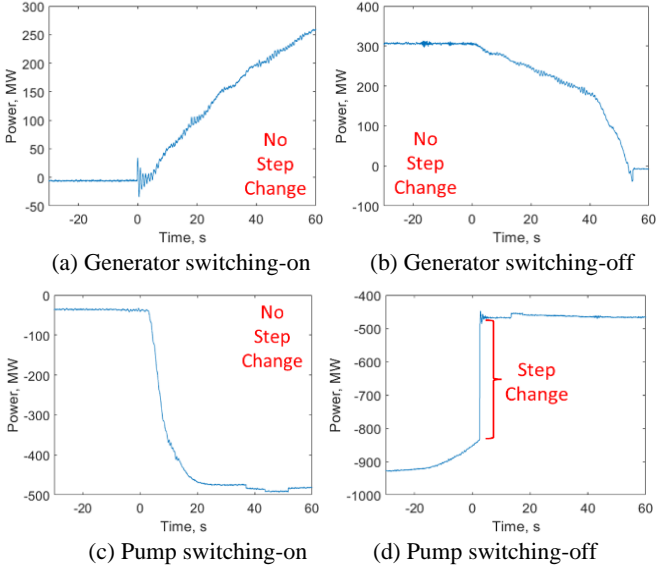

(a) Generator switching-on    (b) Generator switching-off

(c) Pump switching-on    (d) Pump switching-off

**Fig. 5.** PMU power curves for the four types of events in pumped storage hydropower plants.

## C. Physical Process of Pumped Storage Switching-off in Pumped Storage Hydropower Plant

To ensure reliable inertia estimation from pump switching-off events in PSH plants, their physical process is explored. Fig. 6 depicts the PMU power curves for multiple pump switching-off events observed from Bath County PSH plant and Raccoon Mountain PSH plant. Each event shows an initial gradual MW ramp down, followed by a sudden trip with consistent MW step values across events. The physical process behind pump switching-off confirms this MW change phenomenon: upon receiving the switching-off signal, the guide vane of pump unit gradually closes, causing a gradual MW ramp down until a threshold value is reached. At this point, the breaker switches off, resulting in a sudden MW step change. This physical operation confirms the constant sudden MW step change during the pump switching-off event. The consistency in MW step changes eliminates the need for continuous power monitoring and streaming during inertia estimation once the historical constant event MW is known.

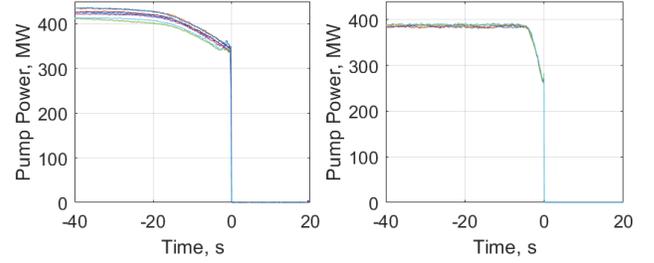

(a) Bath County PSH plant    (b) Raccoon Mountain PSH plant

**Fig. 6.** Observed constant MW step change in confirmed pump switching-off events.

## D. MW Change Distribution of Pumped Storage Hydropower Switching-off Events

To further explore the consistent nature of MW changes during pump switching-off events, PMU data from the Bath County PSH plant was analyzed. Fig. 7 displays a histogram illustrating the MW values recorded during pump switching-off events at this plant in 2021. The histogram reveals a consistent pattern of MW values with minimal deviations. Specifically, the maximum and average absolute deviation ratios from the mean MW are 5.12% and 1.36%, respectively, validating the constant feature of MW step change.

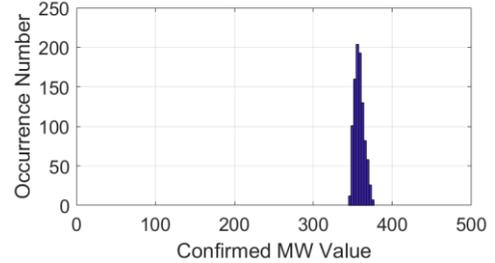

**Fig. 7.** MW distribution for pump switching-off events in Bath County PSH plant.

To delve deeper into the deviation in MW values illustrated in Fig. 7, the monthly maximum, mean, and minimum values of MW during switching-off events were further analyzed, as depicted in Fig. 8. This analysis reveals light variance in pump switching-off MW values across different months, suggesting the absence of any notable seasonal pattern. Hence, the consistent MW values establish a dependable foundation for inertia estimation using pump switching-off events.

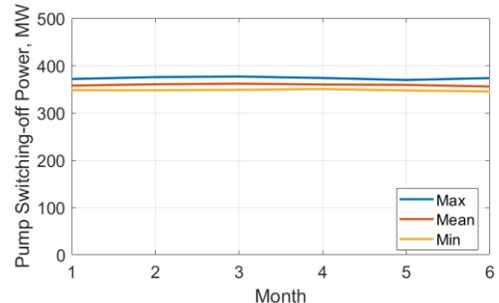

**Fig. 8.** Monthly maximum, mean, and minimum MW change values for switching-off events at Bath County PSH plant.

## E. Occurrence of Confirmed Pump Switching-off Events

The frequency of pump switching-off events is crucial for ensuring an adequate number of inertia estimation points. The daily event count for the Bath County PSH plant is shown in Fig. 9. On average, the Bath County PSH plant experienced 3.6 pump switching-off events per day, with a maximum of 12 events recorded on a single day. In a power system with multiple PSH plants, the number of pump switching-off events would increase, providing more points for inertia estimation. Thus, for a power system with at least one PSH plant, the proposed method mostly can support daily inertia monitoring.

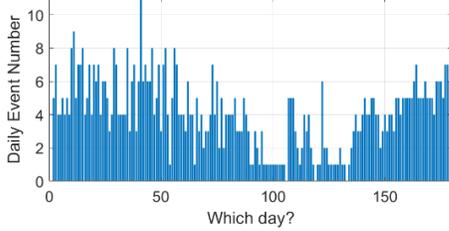

**Fig. 9.** Daily occurrence of switching-off events from a pumped storage hydropower plant.

## F. Comparing Pump Event Time with Daily Lowest Inertia Time

Low inertia can lead to frequency stability issues, so it is beneficial for estimated inertia values to reflect the lowest system inertia each day. Fig. 10(a) illustrates pump switching-off times from the Bath County PSH plant, with each point representing one pump switching-off event, typically occurring between 3 AM and 6 AM. Fig. 10(b) shows the daily lowest inertia times based on U.S. EI system data from the North American Electric Reliability Corporation (NERC), with each point representing the daily lowest inertia, typically occurring from 1 AM to 5 AM. Comparing these time periods reveals that the proposed inertia estimation can provide data points close to the daily lowest inertia, which greatly helps in understanding low inertia situations.

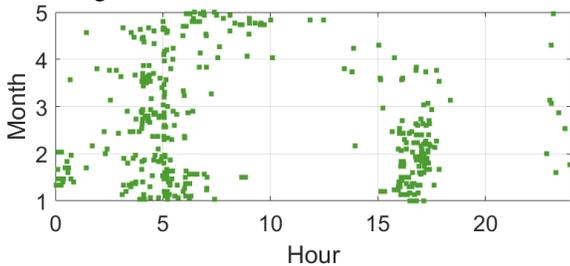

(a) Pump switching-off event time

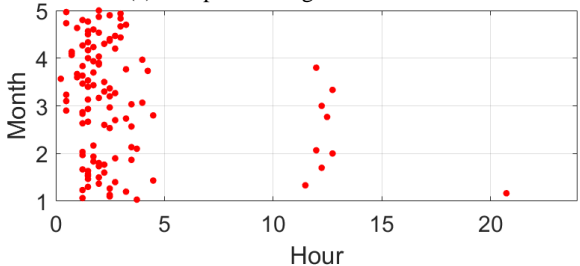

(b) Daily lowest inertia time

**Fig. 10.** Comparison between pump switching-off time and daily lowest inertia time.

## IV. ONLINE EVENT DETECTION TRIGGER DESIGN TO CAPTURE PUMP SWITCHING-OFF EVENTS

Real-time capture of pump switching-off events is crucial for the proposed inertia estimation. An online trigger was designed to detect these events, leveraging three key features of pump switching-off events: local frequency spikes, local voltage step changes, and system frequency deviation. The detection trigger comprises two detection stages: First step uses the simple and robust criteria to quickly and comprehensively the pump switching-off events; Second step uses fine rules to filter out the false alarm, such as the oscillation case. This two-step design based on local features ensures the accuracy of the detection accuracy. The framework of the designed trigger is shown in Fig. 11.

The FNET/GridEye system at ORNL and UTK is a unique monitoring network that provides continuous and real-time independent observations of the entire North American electrical grid [35]. It measures system status including frequency, voltage, and phase angle. Details about the FNET/GridEye measurement system can be found in [36].

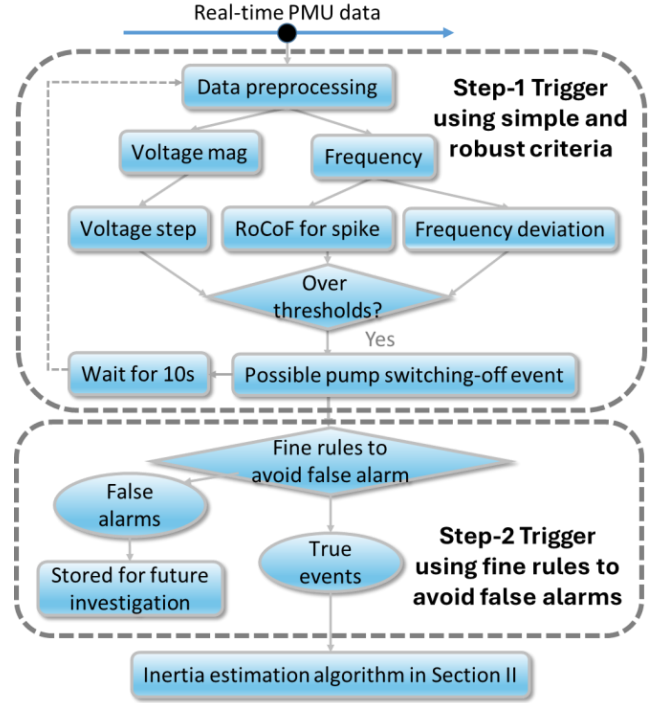

**Fig. 11.** Flowchart for the designed trigger for pump switching-off detection.

## A. PMU Local Measurement Feature for Pumped Storage Switching-off Event

Based on data from FNET/GridEye, local frequency spikes and voltage step changes were noted during pump switching-off events. These features are unique to the local units and cannot be detected by electrically distant units. Fig. 12 illustrates the local frequency spike and voltage step of a confirmed pump switching-off event at the Bath County PSH plant. The FDR located in Roanoke, electrically closer to Bath County, captured the frequency spike and voltage step, while the FDR in Blacksburg (an electrically distant unit) did not.

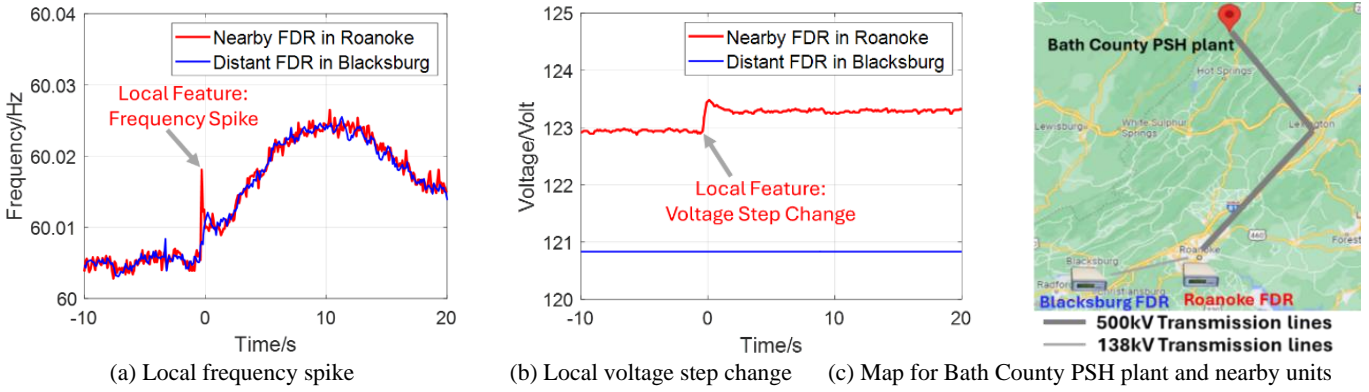

(a) Local frequency spike     (b) Local voltage step change     (c) Map for Bath County PSH plant and nearby units

**Fig. 12.** Measurement comparison between distant and nearby FDRs during pump switching-off events.

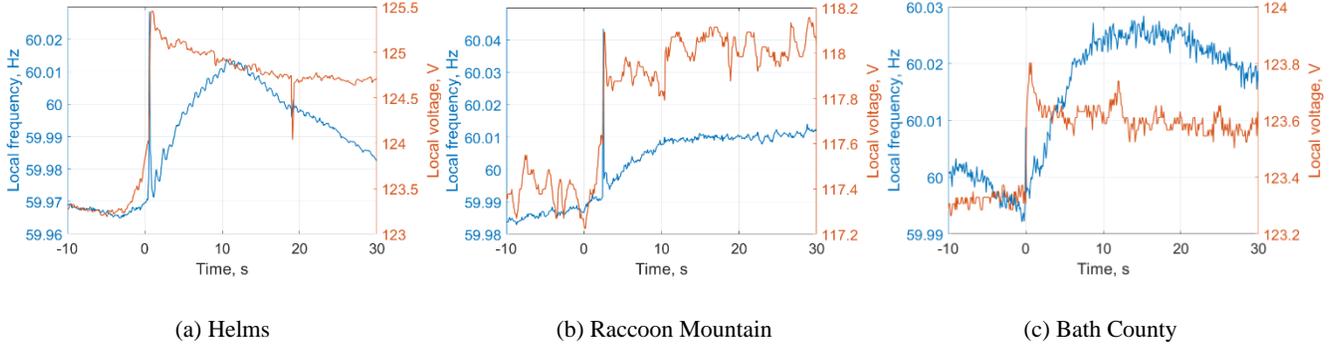

(a) Helms       (b) Raccoon Mountain       (c) Bath County

**Fig. 13.** Local features of frequency spike and voltage step from Helms, Raccoon Mountain, and Bath County PSH plants.

Additionally, these local features during pump switching-off events are widespread across all observed PSH plants. Fig. 13 shows these local features from Helms, Raccoon Mountain, and Bath County PSH plants. These findings ensure that features of frequency spike and voltage step detected by a unit can be attributed to nearby PSH plants, increasing the accuracy of the trigger to capture real-time pump switching-off events.

### B. Real-Time Event Trigger Design to Detect Pump Switching-Off Event

Before estimating inertia, it is essential to detect real-time pump switching-off events. Traditional event triggers rely on significant frequency deviations caused by sudden load or generation changes; however, this method often lacks accuracy and location information. Due to the unique local characteristics of frequency spikes and voltage step changes during pump switching-off events, a novel detection trigger has been designed to accurately capture these real-time events using local features. To enhance detection accuracy, additional refined rules are implemented to filter out false alarms. The two-step trigger process consists of two stages: the first stage swiftly detects events using robust and fast rules, while the second stage applies more refined criteria to eliminate false alarms that pass through the initial detection stage. The detailed process is shown in Fig. 11 and described as follows.

Initially, raw data from FDRs within a moving window undergoes preprocessing. Local voltage step extraction is performed based on local voltage data, and the RoCoF is calculated using short window data to detect local frequency spikes. Additionally, the largest frequency deviation is calculated. These values are then compared with predefined thresholds based on historically confirmed pump switching-off events. If the voltage step, short-window RoCoF, and frequency deviation exceed their predefined thresholds, the trigger records the event time and transmits the possible event data to the Step-2 trigger. To prevent multiple detections of a single event, the trigger waits for a short period before detecting subsequent events. In the Step-2 trigger, additional fine rules are adopted to filter out false alarms, such as those caused by oscillations. These fine rules can also be updated during the application by adding rules to filter out newly noticed false alarms.

To assess accuracy of the designed trigger, a multi-day test was conducted for the Helms PSH plant. During the testing period, 52 confirmed pump switching-off events occurred. Using the Step-1 trigger in Fig. 11, a total of 53 possible pump switching-off events were detected, as depicted in Fig. 14.

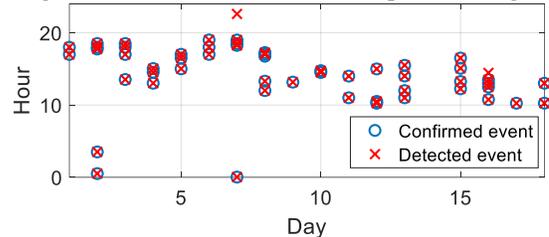

**Fig. 14.** Verification of possible detected events by confirmed events.

Upon comparison with confirmed events, 51 events were validated as genuine. In detail, one confirmed event was missed in detection, resulting in a missing rate of 1.92% (1 out of 52) among confirmed events. Additionally, the Step-1 trigger generated two false alarms, indicating a detection accuracy of 96.23% (51 out of 53) for the trigger. These two false alarms were identified as an oscillation case with significant frequency and voltage deviations. This highlights the potential limitation, particularly in scenarios involving oscillation conditions that may lead to false alarms. To mitigate this issue, a Step-2 trigger is designed and used to filter out such false alarms effectively. Using this improved two-step trigger, testing results show 100% detection accuracy with no false alarms during the testing period.

The missed detection was due to a monitor malfunction that resulted in the absence of frequency and voltage data. During this event, no monitor data was available, leading to its undetected status. This highlights a limitation of relying on a single monitor for event detection. To address this, a multi-monitor-based detection approach has been proposed. This system uses data from multiple local monitors, recording events when any monitor detects one. This approach enhances the sensitivity and reliability of the monitoring system by reducing the likelihood of missed events due to missing data or other single monitor limitations.

Utilizing the designed pump switching-off trigger on FNET/GridEye data, pump switching-off events from Helms, Raccoon Mountain, and Bath County PSH plants were successfully identified in the application.

## V. SIMULATION VERIFICATION OF THE PROPOSED METHOD ON U.S. EASTERN INTERCONNECTION MODEL

To verify the accuracy of the proposed method for estimating system inertia based on pump events, we used the U.S. EI Multiregional Modeling Working Group (MMWG) model with over 60,000 buses. The Bath County PSH plant was selected as the pump switching-off event source for inertia estimation. Random frequency fluctuations were introduced by adding random load changes, creating simulation cases with pre-event RoCoF. A total of 132 cases were generated with varying pre-event and post-event RoCoF: 66 with upward pre-event RoCoF and 66 with downward pre-event RoCoF.

Using the improved RoCoF calculation in (3) and the event MW setting in the simulation, system inertia was estimated for each case. For comparison, inertia was also estimated using the traditional RoCoF-based method, which relies on post-event RoCoF. Because the true inertia from the simulation model can be directly extracted, the error rates for each case, with and without using the improved RoCoF calculation, are shown in Fig. 15. The results clearly demonstrate that the improved RoCoF calculation (red dots) significantly lowers the inertia estimation error rates compared to the traditional method (blue dots). The mean absolute error rate was reduced from 65% to 5% for cases with downward pre-event RoCoF and from 16% to 3% for cases with upward pre-event RoCoF. This proves that the proposed RoCoF calculation method improves the accuracy of RoCoF measurements, resulting in more accurate inertia estimation.

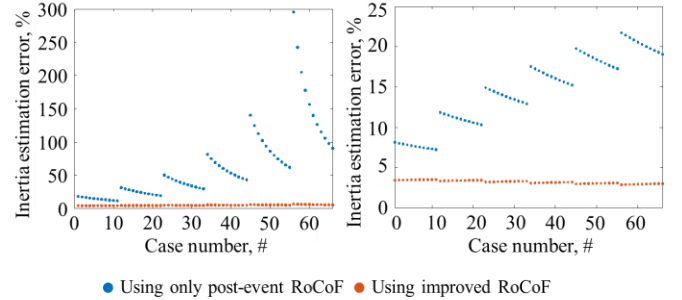

(a) Cases with downward pre-event RoCoF    (b) Cases with upward pre-event RoCoF

• Using only post-event RoCoF    • Using improved RoCoF

**Fig. 15.** Inertia estimation error comparison in Eastern Interconnection System.

## VI. APPLICATION AND FIELD VALIDATION OF THE PROPOSED METHOD ON U.S. WESTERN INTERCONNECTION SYSTEM

The proposed method is applied in the U.S. Western Interconnection System using the Helms PSH. The field data further validated the accuracy of the proposed inertia estimation.

By analyzing the historical MW values of pump switching-off events at the Helms PSH plant, it is confirmed that the MW step changes still remain relatively constant. The average MW step change is approximately 310 MW, with a maximum and minimum deviation of only 7% from the average MW value. This consistency in MW values facilitates accurate inertia estimation for U.S. Western Interconnection, utilizing the average MW value of 310 MW and eliminating the need for real-time MW confirmation.

To validate the accuracy of inertia estimation based on pump switching-off events, confirmed cases from the Helms Pump PSH plant were utilized. The event RoCoF was calculated using FNET/GridEye frequency data. Using the historical event MW and RoCoF values, the inertia for each case was estimated. The accuracy of the estimated inertia was then compared with NERC-reported inertia. Since different RoCoF window sizes can influence RoCoF calculation and affect inertia accuracy, error rates of inertia estimation under various RoCoF window sizes were computed and summarized in Table II. A 0.3s RoCoF window size yielded the most optimal performance for inertia estimation, with an average absolute error rate of 9.9%. This validation demonstrates that the inertia estimation method based on pump switching-off events exhibits satisfactory performance.

TABLE II
PERFORMANCE OF THE INERTIA ESTIMATION IN U.S. WESTERN INTERCONNECTION SYSTEM

| Performance index | RoCoF window size | | | | | | | | |
|---|---|---|---|---|---|---|---|---|---|
| | 0.1s | 0.2s | 0.3s | 0.4s | 0.5s | 1s | 2s | 3s | 4s |
| Median absolute error rate | 28.1% | 8.9% | 8.25% | 8.3% | 13.22% | 29.8% | 52.8% | 68.9% | 82.7% |
| Average absolute error rate | 28.7% | 10.6% | 9.9% | 10.8% | 13.6% | 28.6% | 52.3% | 68.3% | 80.2% |
| Max absolute error rate | 55.0% | 26.8% | 26.1% | 33.2% | 35.8% | 56.7% | 84.2% | 103.7% | 109.63% |

# V. Conclusion

In response to the pressing need for accurate inertia estimation in power systems, this paper introduces a novel practical approach: inertia monitoring based on pump switching-off events. The paper outlines a detailed practice for RoCoF calculation, trigger design, and inertia monitoring, contributing to the following key advancements:

1) A novel inertia estimation method based on PSH switching-off events is proposed. Through comprehensive analysis using field data from PSH plants, the proposed method is promising for grid inertia estimation due to their consistent MW change and frequent occurrence.

2) An accurate inertia estimation algorithm is provided using improved RoCoF, which excludes pre-fault RoCoF from background random frequency fluctuations in practice.

3) A real-time event detection trigger is designed to capture pump switching-off events, leveraging the findings of local frequency spikes, local voltage steps, and system frequency features during these occurrences.

4) The proposed method is validated using the U.S. EI model and implemented in the U.S. Western Interconnection system. Both simulation and field results demonstrate the satisfactory performance of the proposed method.

Despite numerous challenges associated with practical inertia estimation using measurements, this paper presents a novel practical method that achieves high accuracy without additional equipment costs and is easily applicable. The success of inertia estimation using pump switching-off events holds promise for a future carbon-free grid, empowering operators to make informed decisions for operating within high renewable-penetrated power systems with low inertia.